\documentclass[12pt,a4paper]{article}
\pdfoutput=1 % if your are submitting a pdflatex (i.e. if you have
\usepackage{graphicx,amssymb,amsmath,xcolor,float}
\pdfoutput=1
\usepackage{jcappub}
\usepackage[normalem]{ulem}
\graphicspath{{figs/}}

\newcommand{\figw}[2]{\includegraphics[width=#1\textwidth]{#2}}

\newcommand{\black}{\color{black}}
\usepackage[T1]{fontenc} % if needed

\begin{document}
\begin{flushright}
INR-TH-2020-026
\end{flushright}

\title{Constraining superheavy decaying dark matter with directional ultra-high energy gamma-ray limits}

\author[a,b]{O.~Kalashev,}
\author[a,c]{ M.~Kuznetsov,}
\author[a]{Y.~Zhezher}

\affiliation[a]{Institute for Nuclear Research of the Russian Academy of
	Sciences, Moscow, 117312, Russia}
\affiliation[b]{Moscow Institute for Physics and Technology, 9 Institutskiy per., Dolgoprudny, Moscow Region, 141701 Russia}
\affiliation[c]{Service de Physique Th\'eorique, Universit\'e Libre de Bruxelles, Boulevard du Triomphe, CP225, 1050 Brussels, Belgium}

\emailAdd{kalashev@inr.ac.ru}
\emailAdd{mkuzn@inr.ac.ru}
\emailAdd{zhezher.yana@physics.msu.ru}

\abstract{
We present constraints on the lifetime of the superheavy decaying dark matter branching to the $q\bar{q}$ channel in the mass range $10^{19} - 10^{25}$ eV based on the directional limits on the ultra-high-energy (UHE) gamma rays from dwarf spheroidal galaxies (dSphs) and the Milky Way (MW) centre obtained by the Pierre Auger Observatory and the Telescope Array experiment. Attenuation effects during the propagation of UHE photons towards Earth are taken into account. The strongest constraints are derived for the MW centre and have an order of $10^{20}$~yr. We conclude that the UHE diffuse gamma-ray limits provide more efficient signature for the superheavy DM search than the directional gamma-ray limits.
}

\maketitle

\section{Introduction}
\label{sec:intro}

Current list of dark matter (DM) candidates includes tens, if not hundreds, of possibilities. Among the candidates are particles which appear in different extensions of the Standard Model, like supersymmetric partners~\cite{Jungman:1995df}, sterile neutrinos~\cite{Boyarsky:2009ix} and axions~\cite{Abbott:1982af}. Alternatively, one may also suggest macroscopic objects, for example primordial black holes~\cite{Frampton:2010sw}, as well as non-particle scenarios of modified gravity~\cite{Milgrom:1983ca}.

For a long time the weakly interacting massive particles, or WIMPs, were considered as a main cold dark matter candidate. The so-called ``WIMP miracle''~\cite{Jungman:1995df,Kane:2008gb}, the relation between the DM relic density and it's annihilation cross-section being in a good agreement with cosmological predictions, has for a long time motivated the searches for WIMPs in a wide range of possible masses and cross-sections.

Experimental limits have tightly squeezed the possible parameter space, yet no evidence for the detection of a WIMP particle have been obtained so far~\cite{Cushman:2013zza}. Modern constraints are almost touching the so-called ``neutrino floor''~\cite{Billard:2013qya}, unavoidable background related to the neutrino-nucleus scattering, adding complications to further extension of possible range of parameters subjected to tests.

Null results of the WIMP searches have drawn attention to the alternative DM scenarios, one of them being the superheavy dark matter, or SHDM. Historically, superheavy particles were suggested to explain the super-GZK cosmic-ray events~\cite{Berezinsky:1997hy,Kuzmin:1997jua}, later evolving into an independent DM candidate.

It is suggested that SHDM is comprised of non-thermal relics with mass of order of $M_{\chi} \gtrsim 10^{10}\ \mbox{GeV}$ and lifetime much larger than the age of the Universe. It is practically impossible to detect an annihilation of the stable SHDM due to the unitarity constraints on its cross-section (however stable SHDM could be probed by the other means see e.g.~\cite{Tenkanen:2019aij}). The case of decaying dark matter can be tested experimentally more easily, and limits on the high-energy particle fluxes from the DM-rich objects lead to constraints on the $\left( M_{\chi}, \tau\right)$ plane.

Dwarf spheroidal galaxies are one of the promising targets to search for the signal of the dark matter decays~\cite{Lake:1990du,PhysRevD.69.123501}. Dwarf spheroidal galaxies are known to have large mass-to-luminosity ratios~\cite{2008A&A...487..921M,2008MNRAS.386..864D,10.1007/978-94-011-2522-2_13} with low or no astrophysical backgrounds thus being dark matter-dominated.
Data from observations of dwarf spheroidal galaxies in the $\gamma$-ray band by different instruments was used to constrain the dark matter parameters. For example, studies were performed with the data from HAWC~\cite{Albert:2017vtb}, Fermi-LAT~\cite{Ackermann:2015zua,Hoof:2018hyn}, HESS~\cite{Abramowski:2014tra}, MAGIC~\cite{Ahnen:2016qkx} and VERITAS~\cite{Archambault:2017wyh,PhysRevD.85.062001} instruments. So far, the whole range of dark matter masses subjected for the analysis spans from $1$~GeV up to $10^5$~GeV. All of the above mentioned studies have only considered dark matter annihilation in the various channels , except for the HAWC~\cite{Albert:2017vtb} and VERITAS~\cite{PhysRevD.85.062001}, where both DM annihilation and decay cased were analyzed. 
Generally, the are following annihilation and decay channels to be  considered: $W^+W^-$, $ZZ$, $b \bar{b}$, $e^+ e^-$, $gg$, $hh$, $\gamma \gamma$, $\mu^+ \mu^-$, $u \bar{u}$, $d \bar{d}$, $s \bar{s}$, $\tau^+ \tau^-$ and $t \bar{t}$. Each instrument covers some of the channels as well as it's certain energy range. 
In case of annihilation, best constraints are obtained for the $\tau^+ \tau^-$ channel by HESS~\cite{Abramowski:2014tra} and MAGIC~\cite{Ahnen:2016qkx} of order of $10^{-24}\ \mbox{cm}^3 \mbox{s}^{-1}$. For the decaying dark matter, best constraints are derived by HESS~\cite{Abramowski:2014tra} also for the $\tau^+ \tau^-$ decay channel. Lower lifetime limits span from $10^{25}$~s to $10^{27}$~s in the DM mass range from $1$~TeV to $100$~TeV.

In the present paper, searches of $\gamma$-signal from dark matter decays in dwarf spheroidal galaxies are for the first time addressed in the UHE regime with the Pierre Auger Observatory (Auger) and the Telescope Array (TA) experiment data. This allows us to enlarge the possible dark matter masses and constrain the SHDM parameters in the higher mass range than in the mentioned studies. Namely, we employ a set of 20 dwarf spheroidal galaxies adopted from~\cite{10.1093/mnras/stv1601} as well as the Galactic Center (GC). Gamma-ray spectra from the dark matter decay to $q\bar{q}$ channel are calculated with the use of numerical code~\cite{Aloisio:2003xj} and then attenuation effects during the propagation towards Earth are taken into account with the TransportCR code~\cite{Kalashev:1999ma, Kalashev:2014xna}, developed for the simulation of ultra-high-energy cosmic rays and electron-photon cascade attenuation. Obtained spectra are compared with the experimental results on the directional UHE gamma-ray limits from Auger~\cite{Aab:2014bha, Aab:2016bpi} and TA~\cite{Abbasi:2020jgq}, thus allowing to derive a lower bound on the DM lifetime as a function of it's mass.

Previously, a number of studies has addressed the indirect constraints on the SHDM lifetimes. For example, in a similar manner the diffuse high-energy and ultra-high-energy $\gamma$-rays were considered as comprised solely of secondary particles from dark matter decays~\cite{Kalashev:2016cre,Murase:2012xs,Esmaili:2015xpa}. Other class of analyses was aimed to explain the astrophysical neutrino flux observed by the IceCube as a result of dark matter decays~\cite{Murase:2015gea,Kachelriess:2018rty}, which also allows one to derive constraints on it's lifetime. In a recent work~\cite{Ishiwata:2019aet} constraints from various messengers were derived for a wide range of dark matter masses in an unified approach.
The present work continues the series of studies where the relative efficiency of various multi-messenger signatures were tested for SHDM search~\cite{Kalashev:2016cre,Kuznetsov:2016fjt,Kalashev:2017ijd}.

The paper is organized as follows: methods to derive constraints on the DM lifetime are presented in the Section~\ref{sec:methods}, the list of dwarf spheroidal galaxies under assumption and the UHE directional $\gamma$-ray data is described in the Section~\ref{sec:data}. Results and discussion are shown in the Section~\ref{sec:results}.

\section{Methods}
\label{sec:methods}

We analyze the dark matter in the mass range $10^{19} - 10^{25}$ eV, relevant for the constraints from the UHE gamma-rays. For each mass, the injection $\gamma$-spectra are calculated for the $q\bar{q}$ channel, where the DM decay into quarks with uniform distribution in flavors is considered. Decay spectra are obtained with the use of numerical code~\cite{Aloisio:2003xj}, based on the phenomenological approach of deriving the parton fragmentation functions evolved from experimentally measured values with the help of the Dokshitzer-Gribov-Lipatov-Altarelli-Parisi (DGLAP) equations.

Initial fragmentation functions are obtained from the charged-hadron production data~\cite{Hirai:2007cx}, derived at the scale $\sim 1\ \mbox{GeV}$. Then they are extrapolated to the range $10^{-5}\leq \frac{2E}{M_{\chi}} \leq 1$, where $M_{\chi}$ is the DM particle mass and $E$ is the energy of a dark matter decay product. After that, photon injection spectra are calculated analytically, see Ref.~\cite{Kalashev:2016cre} for details. For the energies smaller than $x=10^{-5}$, \black DGLAP equations are no longer valid since one should take into account the coherent branching effects. This doesn't allow one to reliably extrapolate the dark matter decay spectra from obtained ones to the lower energies to utilize all the available UHE $\gamma$-ray data for the $\left( M_{\chi}, \tau\right)$ constraints. Yet we have ensured that the decay spectra from the assumed DM mass range are at least partly covered by the available experimental limits, so that they can be used to constrain the DM parameters.

\begin{figure}%[H]
	\centering
	\includegraphics[width=0.48\textwidth]{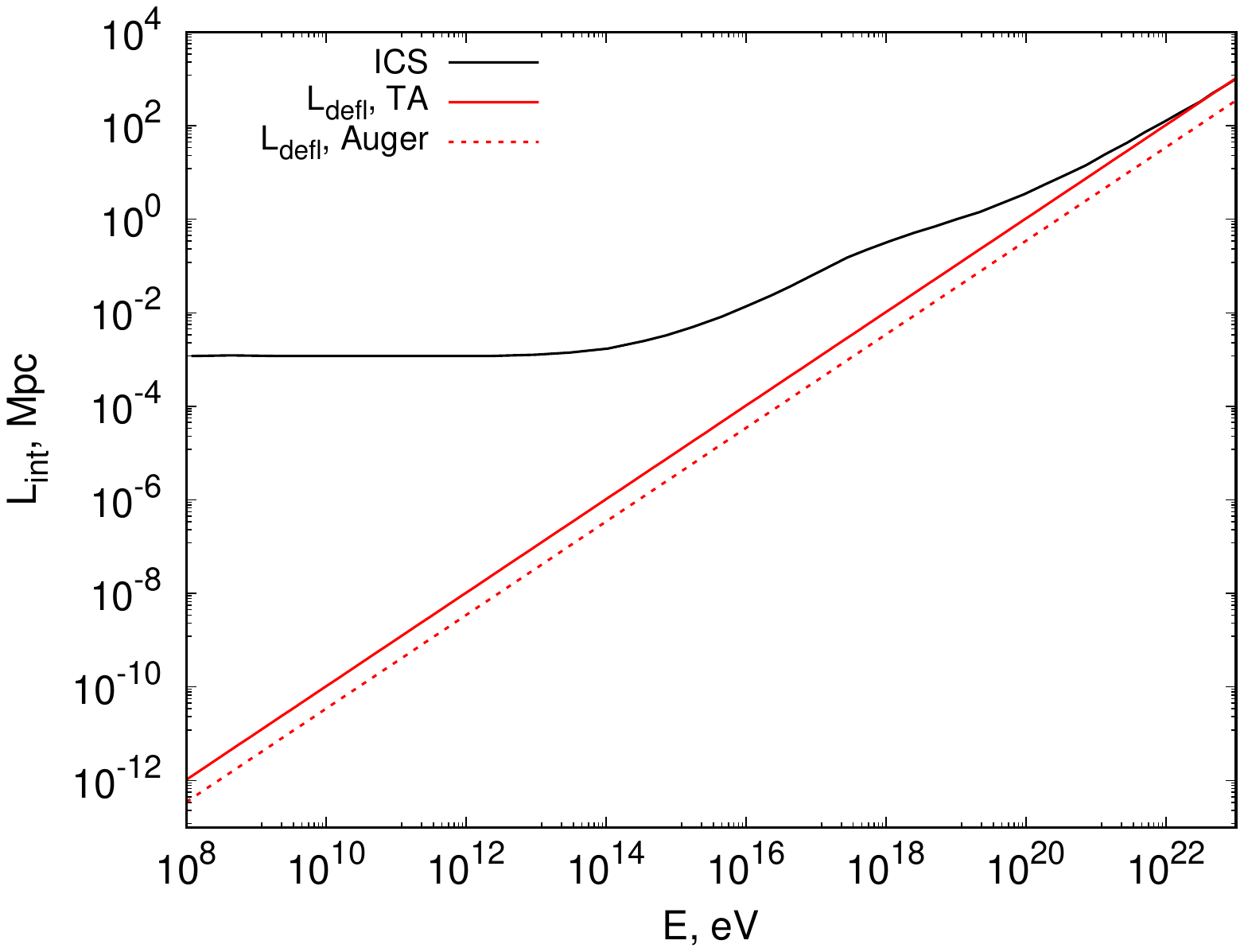}
	\caption{Comparison of the inverse Compton scattering mean free path (solid black line) with the ``deflection distances'' for the TA (solid red line) and Auger (dashed red line).
	}
	\label{fig:icsbend}
\end{figure}

During propagation towards Earth, $\gamma$-rays born in the possible dark matter decays are subjected to attenuation due to interactions with the CMB and EBL that initiate electromagnetic cascades. To take these effects into account, obtained spectra are propagated with the use of the  TransportCR code~\cite{Kalashev:1999ma, Kalashev:2014xna} up to the distance between each dwarf galaxy and Earth, taken from~\cite{10.1093/mnras/stv1601}.

One may notice that secondary $e^+e^-$ from dark matter decays should also contribute to the final $\gamma$-ray signal observed at the Earth, as they initiate electromagnetic cascades due to the inverse Compton scattering (ICS). Electrons and positrons are deflected by Galactic magnetic fields with the approximate curvature radius $R \approx 1.1 \frac{1}{q} \left( \frac{E}{10^{18}\ \mbox{eV}} \right)\ \mbox{kpc}$, and for high enough energies they may bend significantly and leave the instrument's angular resolution pixel faster than they interact and initiate a cascade.

To make sure that electrons and positrons don't contribute to the $\gamma$-ray flux observed at Earth, we have compared the ``deflection distance'', i.e. the distance needed to achieve the deflection angle larger than the angular size of the pixel used in the directional UHE $\gamma$-search with the mean free path for electron/positron of the corresponding energy for the inverse Compton scattering process. The deflection angle is $\Delta \varphi \simeq \frac{l}{R}$, where $l$ is the travel distance and $R$ is the bending radius specified above. If $l$ is smaller than the ICS mean free path, electrons/positrons are assumed to leave the cascade and make no effect on the final $\gamma$-ray flux.

According to~\cite{Aab:2014bha} and~\cite{Abbasi:2020jgq}, in the case of Pierre Auger Observatory, the pixel size $1.0^{\circ}$. In the TA case, pixel size is $3.00^\circ$, $2.92^\circ$, $2.64^\circ$, $2.21^\circ$ and $2.06^\circ$ for energies greater than $10^{18.0}$, $10^{18.5}$, $10^{19.0}$, $10^{19.5}$ and $10^{20.0}$~eV, respectively.
Comparing of ``deflection distances'' with the mean free path for the inverse Compton scattering (see e.g.~\cite{Lee:1996fp}) is shown in the Figure~\ref{fig:icsbend}. The ICS mean free path is shown with solid black line, ``deflection distance'' required to leave TA pixel is shown with solid red line, and for the Auger case -- with dashed red line. For the TA case, we conservatively assume the pixel size of $3.00^\circ$. One may conclude that for all energies considered, electrons/positrons leave the pixel area faster than they upscatter background photons, thus one can neglect contribution of $e^+e^-$ decay products and secondary $e^+e^-$ produced by interaction of photons with CMB for the directional flux calculation.

\begin{figure}%[H]
%	\centering
	\figw{.49}{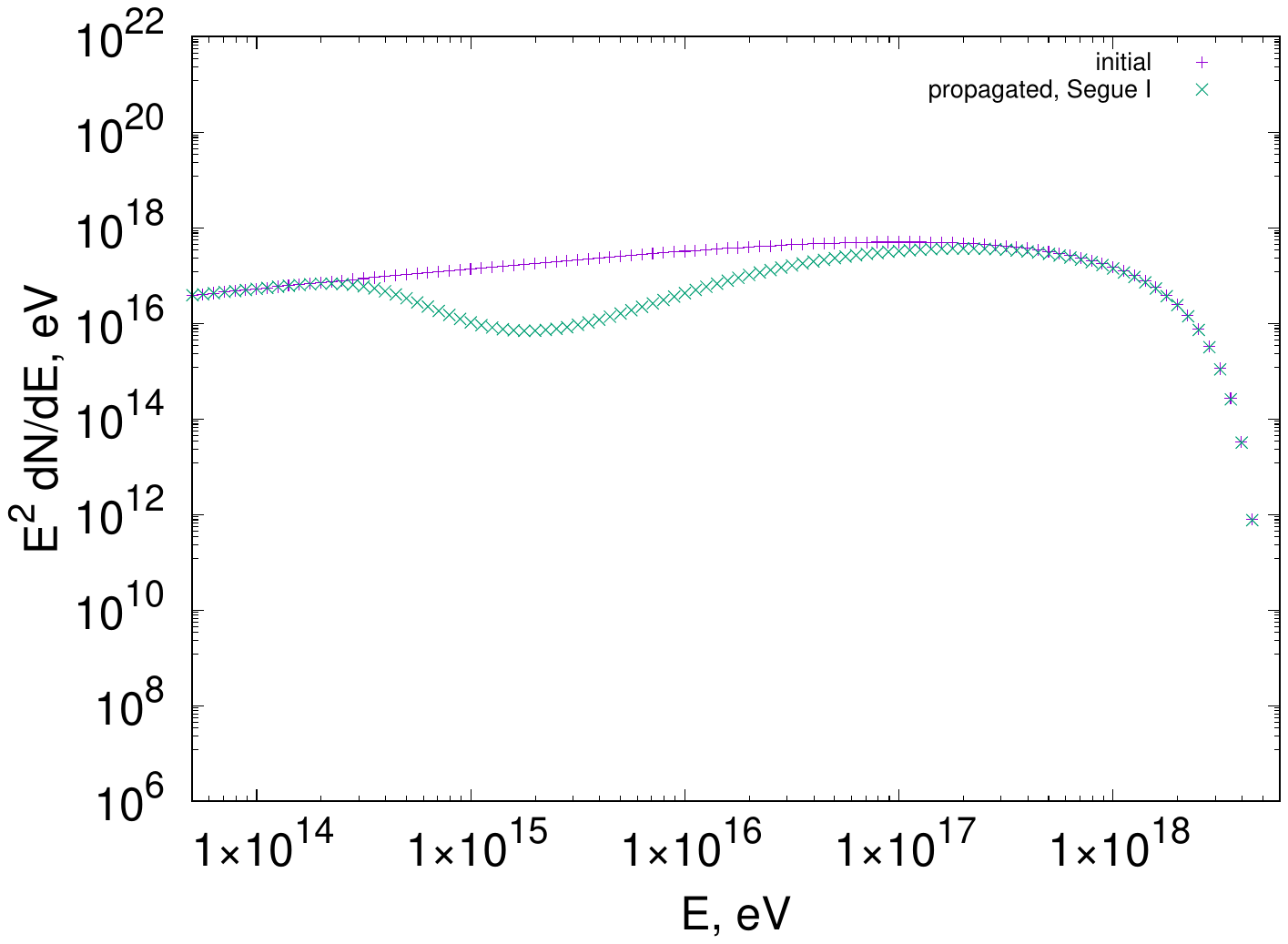}
	\figw{.49}{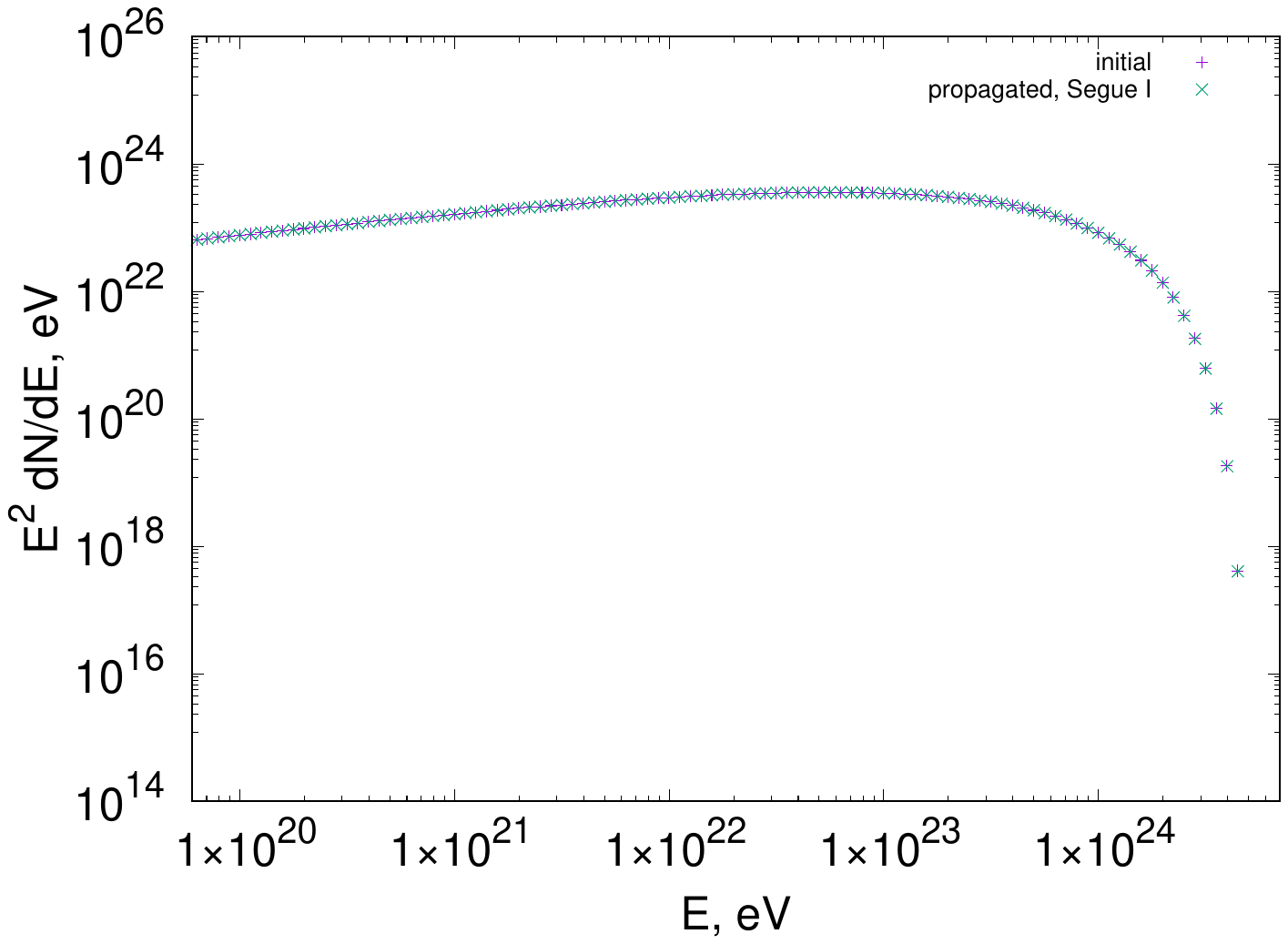}
	\caption{Comparison of initial injection (purple) and propagated (green) photon spectra for the Segue I dwarf spheroidal galaxy, located at the distance $d = 23\ \mbox{kpc}$. Left: DM mass $M_{\chi} = 10^{10}\ \mbox{GeV}$, right: DM mass $M_{\chi} = 10^{16}\ \mbox{GeV}$.
	}
	\label{fig:spectrumseguei}
\end{figure}

\begin{figure}%[H]
	\centering
	\figw{.49}{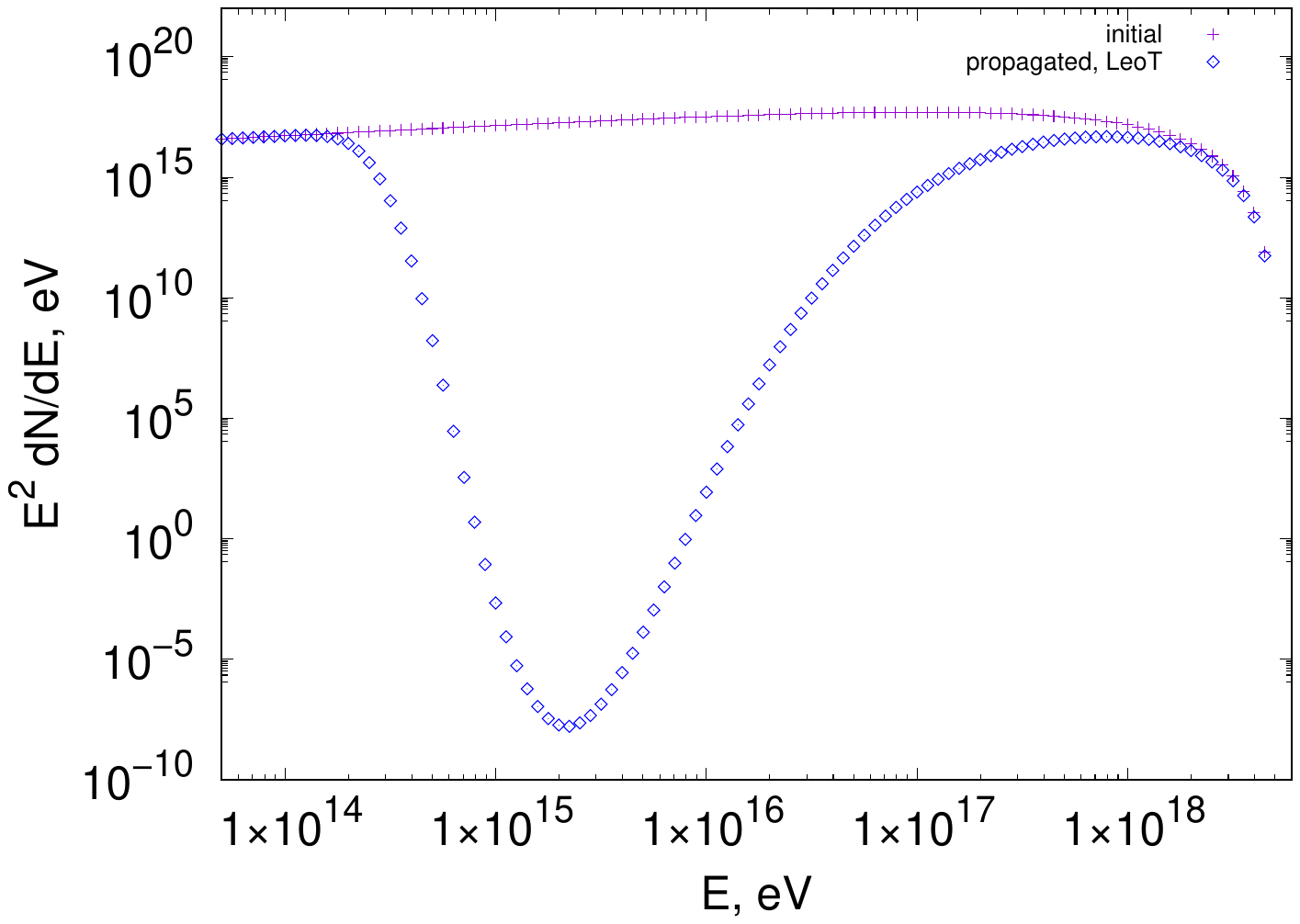}
	\figw{.49}{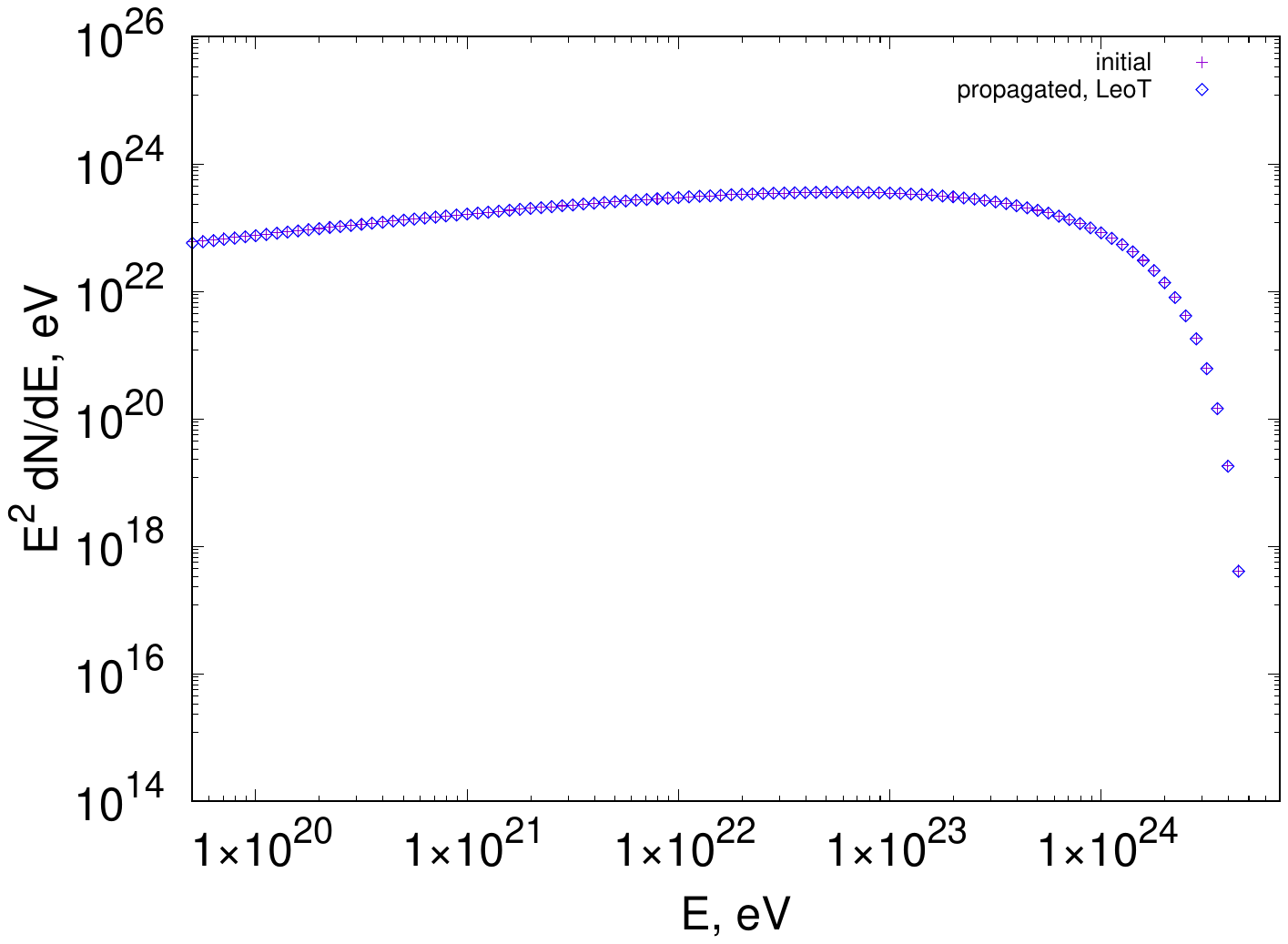}
	\caption{Comparison of initial injection (purple) and propagated (blue) photon spectra for the Leo T dwarf spheroidal galaxy, located at the distance $d = 407\ \mbox{kpc}$. Left: DM mass $M_{\chi} = 10^{10}\ \mbox{GeV}$, right: DM mass $M_{\chi} = 10^{16}\ \mbox{GeV}$.}
	\label{fig:spectrumleot}
\end{figure}

Examples of the propagated $\gamma$-spectra are shown in the Figures~\ref{fig:spectrumseguei} and~\ref{fig:spectrumleot} for the two extreme cases of galaxy location, the closest to Earth, Segue I, located at $23\ \mbox{kpc}$ distance and the furthest, LeoT, located at $407\ \mbox{kpc}$ distance and also for two DM masses: $M_{\chi} = 10^{10}\ \mbox{GeV}$ (left) and $M_{\chi} = 10^{16}\ \mbox{GeV}$ (right). One may see, that the further dwarf galaxy is located, the more attenuation affects the spectra at the lower energies, while for the higher energies the propagation effects are negligible.

After deriving the propagated spectra, one may finally calculate the $\gamma$-ray spectra expected at Earth from a specific dwarf spheroidal galaxy. Following the usual D-factor approach~\cite{Bergstrom:1997fj}, we arrive at the following equation:
\begin{equation} \label{eq:decayflux}
\frac{dF}{dE}_{decay} = \frac{1}{4\pi \tau M_{\chi}}\frac{dN_{\gamma}}{dE}D\enspace,
\end{equation}
%\noindent
where the astrophysical D-factor depends on the actual dark matter distribution in the given dwarf spheroidal galaxy and the distance to it:

\begin{equation} \label{eq:dfactor}
D=\int_{\rm FOV} d\Omega \int_0^{x_{\rm source}} dx \rho (r(\theta,x))\enspace,
\end{equation}
\noindent where $\rho  (r(\theta,x))$ is the distribution of dark matter in a dwarf spheroidal galaxy, the specific profile chosen for the current analysis is described below in the Section~\ref{subsec:dSph}; $r$ is the distance from the Earth to a point within the source, $x$ is the distance along the line of sight, $\theta$ is the angle between the center of the source and the line of sight and FOV denotes the area of the experiment's pixel.

The kinematic studies~\cite{Geringer-Sameth:2014yza} allow one to determine the angular size of a given dwarf spheroidal galaxy. Usually it is calculated as either the half-light radius, radius at which half of the total light of a galaxy is emitted, or for example, the angular distance from the center of a dSph to the outermost member star. Taking into account that dark matter halos spatial extent may be orders of magnitude larger than that of the luminous matter in galaxies, as an upper estimate we calculate D-factors integrating over a size of a pixel corresponding to the angular resolution with respect to the photon primaries of either Auger or TA. The latter is at least couple of times larger than the dSph radius estimates based on the distributions of the luminous matter thus allowing us to take into account possible signals from the ``tails'' of the DM distribution in dwarf spheroidal galaxies.
The decay flux depends on the inverse of the DM mass and lifetime, and for a given DM mass and dwarf galaxy, the constraint on the lifetime may be obtained by normalizing the expected decay $\gamma$-ray flux to the experimental one.

One should also take into consideration, that HE $\gamma$-ray signal from dark matter decays from a given dwarf galaxy is complemented by the contribution from the Milky Way (MW) DM halo from the same direction. One may evaluate the fraction of the UHE $\gamma$-ray flux from the diffuse DM component, assuming the regular Navarro, Frenk \& White (NFW) profile~\cite{Navarro:1995iw} for the dark matter distribution. 
In this case, one can't neglect the contribution from the cascading secondary $e^+e^-$ to the observed $\gamma$-ray flux. For a conservative estimate, we assume the rectilinear propagation of electrons and positrons in the absence of magnetic fields. Together with the flux from propagating secondary $\gamma$-rays, this allows one to calculate the Milky Way DM halo contribution for the direction to the each dwarf spheroidal galaxy. Depending on the proximity to the Galactic Center (GC), MW contribution appears to be of the same order or up to $10000$ times larger than the actual DM signal from a dwarf galaxy itself, even taking into account upper estimate of a dSph size discussed above. The MW contribution appears to be of the same order for Coma, Ursa Major~II and Leo~II dwarf spheroidal galaxies, while for the Hercules galaxy which direction is close to the Galactic Center the MW contribution supersede galaxy's own contribution significantly. Thus the MW DM halo contribution is sumed up with the $\gamma$-ray flux predicted from the dark matter decays in a dwarf galaxy itself and compared with the experimental limits. While mainly the signal from a dSph is considerably smaller than the MW contribution, accounting for both can only strengthens the derived decaying dark matter lifetime constraints. 

The estimation of the HE $\gamma$-ray contribution from the Milky Way DM halo also depends on the choice of the DM density profile. In the present study we have adopted the NFW profile, while other options, such as Einasto~\cite{Graham:2005xx,Navarro:2008kc} or Burkert~\cite{Burkert:1995yz} profiles are also widely used. 

In the same manner as for the dSphs, predicted signal from the dark matter decays in the Milky Way halo depends on the D-factors, where the choice of the DM density profile is enclosed. In~\cite{Cirelli:2010xx}, D-factors were calculated for a variety of DM profiles as a function of an angular distance to the Galactic Center.

To estimate the dependence of the results of the present paper on the DM density profile choice, we compare the D-factors for the case of the dSph closest to the Galactic Center, namely, Hercules. This will give us the upper bound on the systematic shift of the DM lifetime constraints introduced by the DM density profiles since difference between them becomes substantial in the inner halo region, $\lesssim 30^{\circ}$ from the GC.

For the case of Hercules, difference between the NFW, Einasto and Burkert D-factors is $2-4 \%$, which allows us to conclude that the results of the present analysis are practically independent on the choice of the MW DM denisty profile.

\section{Data set}
\label{sec:data}

\subsection{Dwarf spheroidal galaxies}
\label{subsec:dSph}

We employ a set of 20 dwarf spheroidal galaxies, adopted from~\cite{10.1093/mnras/stv1601}. For each dSph, it is necessary to calculate the astrophysical D-factor, which depends on the distribution of dark matter in it.

Universally, for the dark matter profile we adopt the functional form introduced by H.~Zhao~\cite{Zhao:1995cp} to generalize the Hernquist~\cite{1990ApJ...356..359H} profile:

\begin{equation}
\rho(r) = \frac{\rho_s}{\left( r / r_s \right)^\gamma   \left[1 + \left( r / r_s \right)^\alpha \right] ^{ ( \beta -\gamma) / \alpha }}.
\label{eq:rho}
\end{equation}

Parameters $\rho_s$, $r_s$, $\alpha$, $\beta$ and $\gamma$ are measured experimentally~\cite{Geringer-Sameth:2014yza} from available stellar-kinematic data, allowing to perform direct integration and obtain D-factos as shown in the Equation~\ref{eq:dfactor} for each dwarf galaxy independently. Also, following~\cite{Geringer-Sameth:2014yza}, we do not consider Willman~I in the present study as an object with non-equilibrium kinematics~\cite{Willman_2011}.

Full list of analyzed dwarf spheroidal galaxies is given in the Table~\ref{tab:dsph}. Calculated D-factors are shown for the experiment which can observe the given dSph and already include multiplication by the solid angle, which corresponds to the pixel size chosen by either TA or Auger. We also employ a MW GC with a NFW DM profile as a separate source of Auger pixel angular size.

\subsection{UHE directional gamma-ray limits}
\label{subsec:gamma}

In the current analysis, we employ directional limits on the ultra-high-energy gamma rays derived by the Pierre Auger Observatory~\cite{Aab:2014bha, Aab:2016bpi} and the Telescope Array experiment~\cite{Abbasi:2020jgq}.

Auger limits are derived for the declination from $-85^{\circ}$ to $+20^{\circ}$ in the energy range from $10^{17.3}$ eV to $10^{18.5}$ eV, based on the the sample of hybrid events collected between January 2005 and September 2011. No photon point source has been detected, and an upper limit on the photon flux is available for every direction. These limits are set for a pixel size of $1.0^\circ$. We also use a separate point source limit derived by Auger under the same experimental conditions for the Galactic Centre direction in Ref.~\cite{Aab:2016bpi}.

TA limits are based on the Telescope Array surface detector (SD) data obtained during 9 years of observation, with the range of covered declinations $-15.7^{\circ} \leq \delta \leq +85^{\circ}$. As with the Auger case, photon sources are not detected, and upper limits are derived for the point-source flux of UHE $\gamma$-rays with energies greater than $10^{18.0}$, $10^{18.5}$, $10^{19.0}$, $10^{19.5}$ and $10^{20.0}$~eV with pixel sizes of $3.00^\circ$, $2.92^\circ$, $2.64^\circ$, $2.21^\circ$ and $2.06^\circ$ respectively. For the present study, we employ the TA limits derived in the ``real'' background scenario of the mixed
nuclei corresponding to the observed mean $\ln A$.

\section{Results and discussion}
\label{sec:results}

Since Auger and TA limits are integral in energy and derived for separate energy bands, we employ the following approach: if the source is seen by only one of the instruments, constraints are derived only with it's data, while for sources seen by both experiments, the strongest constraint is chosen as the final result.

\begin{figure}%[H]
	\centering
	\includegraphics[width=0.48\textwidth]{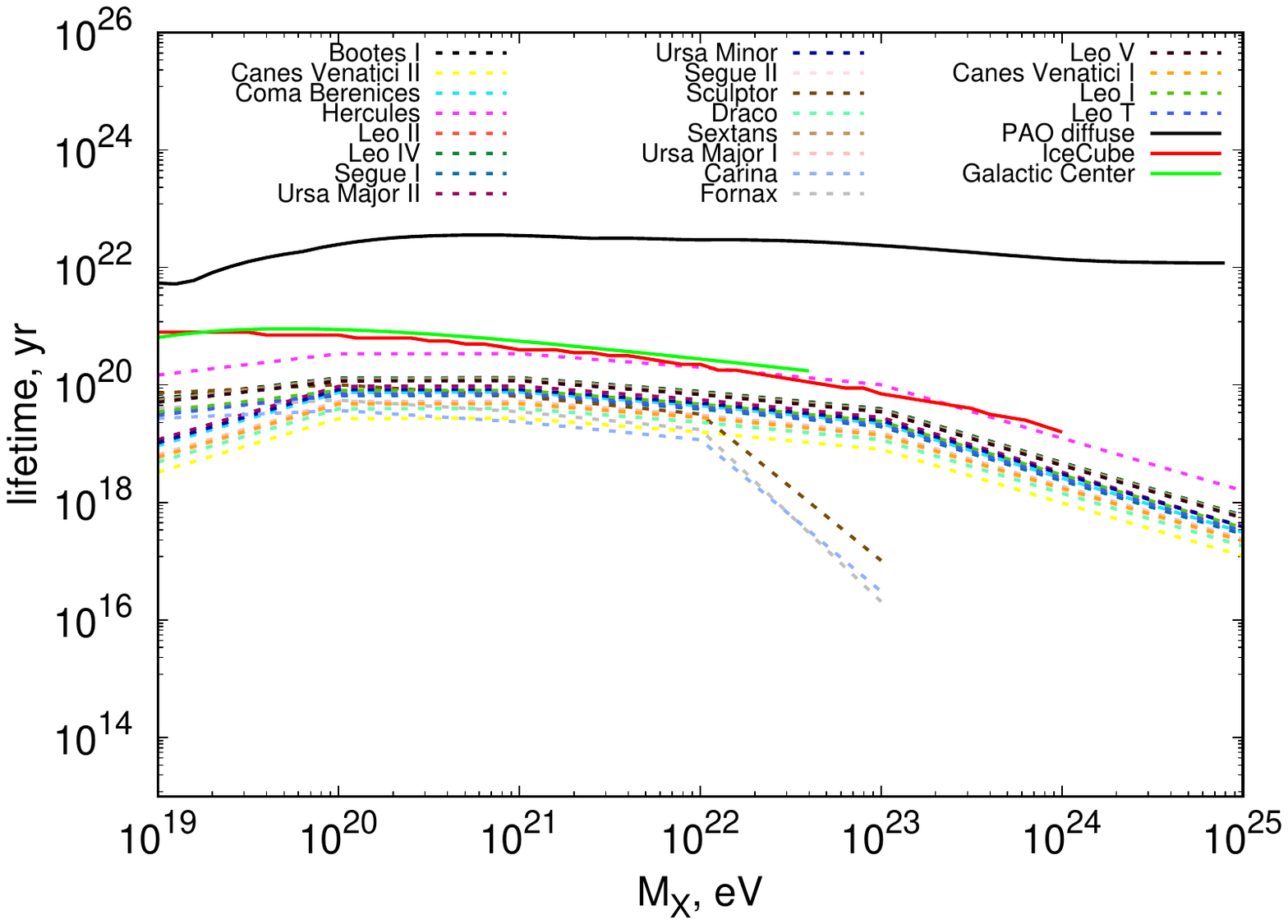}
	\caption{Lower limits on the lifetime of superheavy DM derived for the subset of twenty dwarf spheroidal galaxies (denoted with colors) in the DM mass range DM mass $10^{10}\ \mbox{GeV} \leq M_{\chi} \leq 10^{16}\ \mbox{GeV}$ in comparison with constraints derived from diffuse $\gamma$-ray and neutrino limits from Auger (solid black line) IceCube (solid red line)~\cite{Kachelriess:2018rty}, and Galactic Center (solid green line).}
	\label{fig:limits}
\end{figure}

Final constraints on the lifetime of SHDM are shown in the Figure~\ref{fig:limits}. The constraints derived independently for each dSph and for MW GC. The constraints from the diffuse $\gamma$-ray and neutrino limits of Auger and \mbox{IceCube}~\cite{Kachelriess:2018rty} are also shown for comparison.
One may see that the strongest constraint comes from the MW GC and among the dSphs --- from the Hercules galaxy (that is also because of the large MW contribution). At the same time all the present constraints are looser than those from diffuse $\gamma$-ray limits.
%Predicted flux of $\gamma$-rays from dark matter decays shouldn't excess the limits from any dwarf spheroidal galaxy, thus the strongest constraints derived for the Hercules galaxy also become the final answer of the present analysis.

Let us also discuss possible errors of the dark matter lifetime estimation. One of the sources of uncertainties comes from the accuracy of the D-factor estimation. In the D-factor calculation, we employ the median values of parameters $\rho_s$, $r_s$, $\alpha$, $\beta$ and $\gamma$ in the dark matter density profiles of dwarf spheroidal galaxies. $1 \sigma$ upper and lower values of these parameters correspond to uncertainties in D-factor estimation of a few percent, which leads to the lifetime errors of the same order.

It is possible to consider different dark matter distribution profiles as an alternative to the H.~Zhao profile chosen for the dwarf spheroidal galaxies and the NFW profile chosen for the Milky Way DM halo contribution. It was shown by Bonnivard et al.~\cite{10.1093/mnras/stu2296}, that the different parametrizations -- Zhao–Hernquist or Einasto have negligible impact on the calculated D-factors and their uncertainties. And as was estimated in~\cite{Kalashev:2016cre}, implementation of the Burkert profile instead of the NFW for the Milky Way leads to negligible difference in the predicted fluxes of secondary particles from dark matter decays in the Galactic halo unless we consider sources close to the Galactic Center. For the GC region the NFW profile would yield stronger constraint than cored profiles, therefore our result for the GC should be interpreted as the upper bound of what one could expect.  

Calculated DM decay spectra uncertainties also add up to the uncertainties of the estimated DM lifetime constraints. Mainly, following~\cite{Kalashev:2016cre, Kachelriess:2018rty}, we employ only photons born in the pion decays and neglect the contribution from the kaon decays, which make up to 10 \% of the pion flux.
Electroweak corrections also result in the additional photons, which are produced not in the hadron decays. As it was shown in~\cite{Ciafaloni:2010ti,Cirelli:2010xx}, one may also disregard corresponding photon fluxes as negligible in comparison with the primary one.

In the present paper, the constraints on the superheavy dark matter lifetimes were obtained with the use of the directional UHE $\gamma$-ray limits from the Auger and TA experiments which allowed us to exploit the dwarf spheroidal galaxies and the MW GC as possible sources of DM signal in the mass range $10^{19} - 10^{25}$ eV for the first time. Derived constraints appear to be at least an order of magnitude looser than the ones obtained from the Auger diffuse gamma-ray limits and somewhat looser than those from IceCube neutrino limits~\cite{Kachelriess:2018rty}.
The main reason for this is that in the absence of the actually observed flux the constraining power is determined by two factors. From one hand, it is affected by the effective angular size of the field of view and from the other hand, by the model signal to background ratio in this field of view. Therefore, one would hope that the decrease in the FoV angular size from a full experiment FoV to a point-source pixel size would be overshot by the expected signal to background ratio growth. Our result shows that this does not happen. This implies the relative non-efficiency of the point-source UHE $\gamma$-ray limits as the signature for the SHDM search.

\acknowledgments

The work was supported by the Foundation for the Advancement of Theoretical Physics and Mathematics “BASIS” grant 17-12-205-1.

\pagebreak

%\onecolumngrid

%\twocolumngrid

%\onecolumngrid

\begin{table}
%\centering
\begin{tabular}{|c|c|c|c|}
\hline
Name & Distance ($\mbox{kpc}$) & $D_{Auger}\ \left(\mbox{GeV}/{\mbox{cm}}^2\right)$ & $D_{TA}\ \left(\mbox{GeV}/{\mbox{cm}}^2\right)$ \\
\hline
Segue 1 & $23.0$ &	$4.8 \times 10^{18}$ & $2.0 \times 10^{19}$\\
\hline
Ursa Major II & $30.0$ & $-$ & $5.1 \times 10^{19}$\\
\hline
Segue 2 & $35.0$ & $-$ & $2.2 \times 10^{16}$\\
\hline
Coma Berenices & $44.0$	& $-$ & $7.0 \times 10^{19}$\\
\hline
Ursa Minor & $66.0$ & $-$ & $2.0 \times 10^{17}$\\
\hline
Bootes I & $66.0$ & $2.8 \times 10^{18}$ & $1.5 \times 10^{19}$\\
\hline
Sculptor & $79.0$ & $2.7 \times 10^{17}$ & $-$\\
\hline
Draco & $82.0$ & $-$ & $4.2 \times 10^{18}$\\
\hline
Sextants & $86.0$ &	$6.5 \times 10^{17}$ & $2.8 \times 10^{18}$\\
\hline
Ursa Major & $97.0$ & $-$ & $1.4 \times 10^{18}$\\
\hline
Carina & $101.0$ & $2.1 \times 10^{17}$ & $-$\\
\hline
Hercules & $132.0$ & $2.2 \times 10^{16}$ & $2.3 \times 10^{16}$\\
\hline
Fornax & $138.0$ & $9.9 \times 10^{16}$ & $-$\\
\hline
Leo IV & $160.0$ & $1.32 \times 10^{16}$ & $1.34 \times 10^{16}$\\
\hline
Canes Venatici II & $160.0$	& $-$ & $1.6 \times 10^{19}$\\
\hline
Leo V & $180.0$ & $3.0 \times 10^{17}$ & $3.7 \times 10^{17}$\\
\hline
Leo II & $205.0$ & $-$ & $5.0 \times 10^{16}$\\
\hline
Canes Venatici I & $218.0$ & $-$ & $1.2 \times 10^{17}$\\
\hline
Leo I & $250.0$ & $6.1 \times 10^{17}$ & $9.9 \times 10^{17}$\\
\hline
Leo T & $407.0$ & $6.1 \times 10^{17}$ & $9.9 \times 10^{17}$\\
\hline
\end{tabular}
\caption{\label{tab:dsph} List of dwarf spheroidal galaxies used in the present study to obtain constraints on the dark matter lifetime. Only if the galaxy is in the field of view (FoV) of either Auger or TA, the corresponding D-factor is shown. D-factors already include the solid angle, which corresponds to the size of pixel chosen for the pixelization in the UHE $\gamma$ search by TA and Auger.}
\end{table}

%\twocolumngrid

\pagebreak

\bibliographystyle{JHEP}
\bibliography{dSph}

\end{document}